\documentclass[12pt, a4paper]{article}
\usepackage{latexsym,amsmath,amsfonts,amssymb}
\usepackage{graphicx}
\usepackage{indentfirst}
\usepackage{cite}
\usepackage[cp1251]{inputenc}

\newtheorem{theorem}{Theorem}
\newtheorem{lemma}{Lemma}

\begin{document}

\begin{center}
 {\Large \bf Symmetry analysis and exact solutions of one class of (1+3)-dimensional boundary-value problems of the Stefan type}

\medskip \medskip

{\bf S.\,S. Kovalenko}
\medskip \medskip \medskip
\\
{\it Institute of Mathematics, Ukrainian National Academy of Sciences,
\\
Kyiv, Ukraine}
\\
(e-mail: kovalenko@imath.kiev.ua)

\medskip \medskip \medskip

\end{center}

\begin{abstract}
We present the group classification of one class of (1+3)-dimensional nonlinear boundary-value problems of the Stefan type that simulate the processes of melting and evaporation of metals. The results obtained are used for the construction of the exact solution of one boundary-value problem from the class under study.
\end{abstract}


\medskip

\textbf{UDC} 517.957+512.816.

\newpage

\section{Introduction}

Group-theoretic methods form a contemporary mathematical apparatus widely used in the investigation of mathematical models on the basis of differential (ordinary and partial) equations. By this time, the symmetry properties of many known equations of mechanics, gas dynamics, electrodynamic, thermal physics, quantum physics, etc., have been studied (see \cite{ovs78,fus90,fus89,lah02,blu02} and the references therein). However, in most works, differential equations were investigated without taking into account any initial and boundary conditions. This can be explained by the fact that the widely used boundary conditions (e.g., Dirichlet and Neumann conditions, etc.) are seldom invariant under transformations from the symmetry group of basic differential equations. This, in turn, leads to a low efficiency of group methods in the search for, e.g., exact solutions of boundary-value problems.

One of possible approaches to handling this difficulty is the application of group-theoretic methods to boundary-value problems with moving boundaries (free surfaces). Although these problems are more difficult to analyze than traditional boundary-value problems with fixed boundaries, the application of, say, the classic Lie method to these problems may be much more efficient because the structure of unknown (free) boundaries can depend on invariant variables, which enables one to reduce the boundary-value problem under study to another problem of smaller dimension \cite{che09}. There are known works in which group-theoretic methods are applied to boundary-value problems with moving boundaries in the case of two independent variables (see \cite{che10} and the references therein). However, at present, there are only a few works devoted to the application of group methods to complex multidimensional problems with moving boundaries \cite{che03,puk72,ben82}, though the results obtained in these works testify to the efficiency of this approach.

In the present paper, we consider a broad class of (1+3)-dimensional boundary-value problems of the Stefan type that simulate the processes of melting and evaporation of metals under the action of powerful radiation flows. In Sec.~2, we formulate a theorem that gives an exhausting description of the Lie symmetries of boundary-value problems from the class considered. In Sec.~3, the results of group classification are used for reduction and for the construction of an exact solution of a boundary-value problem from the class considered.

\section{Invariance of One Class of (1+3)-Dimensional Boundary-Value Problems that Simulate the Processes of Melting and Evaporation of Metals}

We formulate a mathematical model of the processes of melting and evaporation of metals under the action of powerful radiation flows. Let $\Omega = \{ \boldsymbol{x} = (x_1, x_2, x_3): x_3 > 0 \}$ be a half-space of the three-dimensional Euclidean space $\mathbb{R}^3$ occupied initially by the solid phase of a metal. At time $t = 0$, a powerful radiation flow $\boldsymbol{Q}(t) = (0, 0, Q(t))$ starts falling on the surface $x = 0$. Here and in what follows, we neglect the initial short-term nonequilibrium stage of the process and study the processes of melting and evaporation in the stage where the three phases (gaseous, liquid, and solid) are present, assuming that this situation arises at time $t \in \mathfrak{T} = (t_{\ast}, + \infty)$, where $t_{\ast}$ is a certain positive real number. Thus, the domain $\Omega(t) = \Omega \times \mathfrak{T}$ consist of the three domains occupied by gaseous, liquid, and solid phases (denote by $\Omega_0(t)$, $\Omega_1(t)$, and $\Omega_2(t)$, respectively) and the two smooth surface $S_1(t, \boldsymbol{x}) = 0$ and $S_2(t, \boldsymbol{x}) = 0$ that separate these phases. In other words, the domain $\Omega(t)$ can be represented as follows:
\[
    \Omega(t) = \Omega_0(t) \cup \Gamma_1(t) \cup \Omega_1(t) \cup \Gamma_2(t) \cup \Omega_2(t),
\]
where
\[
    \Gamma_k(t) = (t, \boldsymbol{x}): S_k(t, \boldsymbol{x}) = 0, \ t \in \mathfrak{T}, \ \boldsymbol{x} \in \Omega, \ \ k = 1, 2,
\]
\[
    \Omega_0(t) = (t, \boldsymbol{x}): S_1(t, \boldsymbol{x}) < 0, \ S_2(t, \boldsymbol{x}) < 0, \ t \in \mathfrak{T}, \ \boldsymbol{x} \in \Omega,
\]
\[
    \Omega_1(t) = (t, \boldsymbol{x}): S_1(t, \boldsymbol{x}) > 0, \ S_2(t, \boldsymbol{x}) < 0, \ t \in \mathfrak{T}, \ \boldsymbol{x} \in \Omega,
\]
\[
    \Omega_2(t) = (t, \boldsymbol{x}): S_1(t, \boldsymbol{x}) > 0, \ S_2(t, \boldsymbol{x}) > 0, \ t \in \mathfrak{T}, \ \boldsymbol{x} \in \Omega.
\]

Then the corresponding class of (1+3)-dimensional nonlinear boundary-value problems of the Stefan type that simulate the processes of melting and evaporation of metals under the action of powerful radiation flows can be represented in the form \cite{ani70,lyu82}
\begin{equation}\label{1}
    \frac{\partial u}{\partial t} = \nabla \, (d_1(u) \nabla u), \ \ (t, \boldsymbol{x}) \in \Omega_1(t),
\end{equation}
\begin{equation}\label{2}
    \frac{\partial v}{\partial t} = \nabla \, (d_2(v) \nabla v), \ \ (t, \boldsymbol{x}) \in \Omega_2(t),
\end{equation}
\begin{equation}\label{3}
    S_1(t, \boldsymbol{x}) = 0: \, d_{1\,v} \frac{\partial u}{\partial \boldsymbol{n}_1} = H_v \boldsymbol{V}_1 \cdot \boldsymbol{n}_1 - \boldsymbol{Q}(t) \cdot \boldsymbol{n}_1, \ \ u = u_v,
\end{equation}
\begin{equation}\label{4}
    S_2(t, \boldsymbol{x}) = 0: \, d_{2\,m} \frac{\partial v}{\partial \boldsymbol{n}_2} = d_{1\,m} \frac{\partial u}{\partial \boldsymbol{n}_2} + H_m \boldsymbol{V}_2 \cdot \boldsymbol{n}_2, \ \ u = u_m, \ \ v = v_m,
\end{equation}
\begin{equation}\label{5}
    |\,\boldsymbol{x}| = + \infty: \, v = v_{\infty},  \ \ t \in \mathfrak{T},
\end{equation}
where $u$ and $v$ are the required temperature fields; $d_1(u)$ and $d_2(v)$ are thermal diffusivities, $d_{1\,v} = d_1(u_v)$, $d_{1\,m} = d_1(u_m)$, $d_{2\,m} = d_2(v_m)$; $H_v$, $H_m$, $u_v$, $u_m$, $v_m$, and $v_{\infty}$ are certain known constants; $\boldsymbol{Q}(t) = (0, 0, Q(t))$ is a heat flow; $S_k(t, \boldsymbol{x}) = 0$, $k = 1, 2$, are the required interfaces of phases; $\boldsymbol{V}_k(t, \boldsymbol{x})$, $k = 1, 2$, are the velocities of motion of interphase boundaries; $\boldsymbol{n}_k(t, \boldsymbol{x})$, $k = 1, 2$, are the unit outward normals to the surfaces $S_k(t, \boldsymbol{x}) = 0$, $k = 1, 2$, respectively;
\[
    \nabla = \left(\frac{\partial}{\partial x_1}, \frac{\partial}{\partial x_2}, \frac{\partial}{\partial x_3} \right);
\]
and the indices $k = 1$ and $k = 2$ denote, respectively, the liquid and the solid phase of the metal.

From the viewpoint of mathematics and physics, we must impose certain additional conditions on functions and constants in the considered class of boundary-value problems. Namely, we assume that all functions of problem \eqref{1}--\eqref{5} are sufficiently smooth and the free surfaces $S_k(t, \boldsymbol{x}) = 0$ satisfy the following conditions:
\[
    \frac{\partial S_k}{\partial t} \neq 0, \ |\nabla S_k| \neq 0, \ \ k = 1, 2,
\]
the heat flow is nonzero, i.e., $Q(t) \neq 0$, and $\boldsymbol{V}_k \cdot \boldsymbol{n}_k \neq 0$, $k = 1, 2$. Finally, the constants $u_v$, $u_m$, $v_m$, and $v_{\infty}$ must satisfy the natural inequalities $u_v \neq u_m$ and $v_m \neq v_{\infty}$.

On the basis of the definition of the Lie invariance of a boundary-value problem with free surfaces proposed in \cite{che10}, using the classic methods of group classification of differential equations \cite{ovs78, lah02, blu02}, we investigate the symmetry properties of the class of boundary-value problems \eqref{1}--\eqref{5}. Lemma~1 and Theorem~1 presented below give an exhaustive description of the Lie symmetries of boundary-value problems from this class.

\begin{lemma}\label{l1}
    The class of boundary-value problems \eqref{1}--\eqref{5} admits the group of equivalence transformations $\widetilde{E}_{\mathrm{eq}}^{\mathrm{BVP}}$:
    \[
        \tilde{x}_1 = \beta \, (x_1 \cos\beta_1 + x_2\sin\beta_1) + \gamma_1,
    \]
    \[
        \tilde{x}_2 = \beta \, (-x_1 \sin\beta_1 + x_2\cos\beta_1) + \gamma_2,
    \]
    \[
        \tilde{x}_3 = \beta _3 + \gamma_3,
    \]
    \[
        \tilde{t} = \alpha t + \gamma_0, \ \ \tilde{u} = \delta_1 u + \gamma_4, \ \ \tilde{v} = \delta_2 v + \gamma_5, \ \ \widetilde{S}_1 = S_1, \ \ \widetilde{S}_2 = S_2,
    \]
    \[
        \tilde{d}_1 = \frac{\beta^2}{\alpha} \, d_1, \ \ \tilde{d}_2 = \frac{\beta^2}{\alpha} \, d_2,
    \]
    \[
        \tilde{d}_{1\,v} = \frac{\beta^2}{\delta_1} \, d_{1\,v}, \ \ \tilde{d}_{1\,m} = \frac{\beta^2}{\delta_1} \, d_{1\,m}, \ \ \tilde{d}_{2\,m} = \frac{\beta^2}{\delta_2} \, d_{2\,m}, \ \ \widetilde{H}_v = \alpha H_v, \ \ \widetilde{H}_m = \alpha H_m,
    \]
    \[
        \tilde{u}_v = \delta_1 u_v + \gamma_4, \ \ \tilde{u}_m = \delta_1 u_m + \gamma_4, \ \ \tilde{v}_m = \delta_2 v_m + \gamma_5, \ \ \widetilde{Q} = \beta Q,
    \]
    where $\alpha$, $\beta$, $\beta_1$, $\gamma_0, \ldots, \gamma_5$, $\delta_1$, and $\delta_2$ are arbitrary real coefficients that satisfy the condition
    \[
        \alpha \beta \delta_1 \delta_2 \neq 0.
    \]
\end{lemma}

\begin{theorem}\label{t1}
    The boundary-value problem \eqref{1}--\eqref{5} with arbitrary given real functions $d_1(u)$, $d_2(v)$, and $Q(t)$ is invariant in the Lie sense with respect to the 4-parameter Lie group generated by the infinitesimal operators
    \[
        P_1 = \frac{\partial}{\partial x_1}, \ \ P_2 = \frac{\partial}{\partial x_2}, \ P_3 = \frac{\partial}{\partial x_3}, \ \ J_{1 \, 2} = x_2 \frac{\partial}{\partial x_1} - x_1 \frac{\partial}{\partial x_2}.
    \]
    The maximal invariance group of problem \eqref{1}--\eqref{5} does not depend on the form of the functions $d_1(u)$ and $d_2(v)$ but depends on the function $Q(t)$. There exist only two boundary-value problems from class \eqref{1}--\eqref{5} (up to equivalence transformations from the group $\widetilde{E}_{\mathrm{eq}}^{\mathrm{BVP}}$) with properly chosen function $Q(t)$ that admit the 5-parameter invariance group generated by $P_1$, $P_2$, $P_3$, $J_{1 \, 2}$, and the operators
    \[
        D = 2t \frac{\partial}{\partial t} + x_1 \frac{\partial}{\partial x_1} + x_2 \frac{\partial}{\partial x_2} + x_3 \frac{\partial}{\partial x_3} \ \ \mbox{for} \ \ Q(t) = \frac{q}{\sqrt t}
    \]
    and
    \[
        P_t = \frac{\partial}{\partial t} \ \ \mbox{for} \ \ Q(t) = q,
    \]
    respectively, where $q$ is an arbitrary real constant.
\end{theorem}

\textbf{\emph{Remark 1.}} The arbitrary constant $q$ in Theorem~1 can be set equal to 1 up to equivalence transformations from the group $\widetilde{E}_{\mathrm{eq}}^{\mathrm{BVP}}$ (recall that $Q(t)$ is a real function).

\textbf{\emph{Remark 2.}} The case $d_1(u) = d_2(v) = \mathrm{const}$ is not investigated in the proof of the theorem because it is not physically substantiated.

\textbf{\emph{Remark 3.}} In the case of one space variable $x_3$, problem \eqref{1}--\eqref{5} is essentially simplified. There are numerous works (see, e.g., \cite{che09,che10,che90,che93}) devoted to the determination of exact solutions and Lie symmetries of this problem.

\section{Symmetry Reduction and an Example of the Construction of an Exact Solution of One Boundary-Value Problem from the Considered Class}

In this section, we use the results obtained in Sec.~2 for the construction of the exact solutions of boundary-value problems from the considered class. Consider the boundary-value problem \eqref{1}--\eqref{5} with heat flow $\boldsymbol{Q}(t) = \boldsymbol{q} \equiv (0, 0, q), \ q = \mathrm{const}$. According to Theorem~1, this problem admits the 5-dimensional invariance algebra $A_5$ with the basis operators
\[
    P_t = \frac{\partial}{\partial t}, \ \ P_1 = \frac{\partial}{\partial x_1}, \ \ P_2 = \frac{\partial}{\partial x_2}, \ P_3 = \frac{\partial}{\partial x_3}, \ \ J_{1 \, 2} = x_2 \frac{\partial}{\partial x_1} - x_1 \frac{\partial}{\partial x_2}.
\]
We use these operators for the reduction of this boundary-value problem to boundary-value problems of lower dimension. For this purpose, we use an optimal system of $s$-dimensional subalgebras ($s \leq 5$) of the algebra $A_5$. Since the algebra $A_5$ can be represented in the form $A_5 = \langle P_1, P_2, J_{1, \, 2} \rangle \oplus \langle P_3 \rangle \oplus \langle P_t \rangle$, using the known Lie--Goursat algorithm for the classification of subalgebras of Lie algebras decomposable into a direct sum \cite{pat75} and the results of classification of subalgebras of low-dimensional real Lie algebras \cite{pat77} we obtain the following complete list of required subalgebras:

subalgebras of dimension 1:
\[
    \langle P_3 \cos\phi + P_t \sin\phi \rangle, \ \ \langle P_1 + \alpha \, (P_3 \cos\phi + P_t \sin\phi) \rangle,
\]
\[
    \langle J_{1\,2} + \beta \, (P_3 \cos\phi + P_t \sin\phi) \rangle;
\]

subalgebras of dimension 2:
\[
    \langle P_3, P_t \rangle, \ \ \langle P_1 + \alpha \, (P_3 \cos\phi + P_t \sin\phi), P_2 \rangle,
\]
\[
    \langle P_1 + \alpha \, (P_3 \cos\phi + P_t \sin\phi), P_3 \sin\phi - P_t \cos\phi \rangle,
\]
\[
     \langle J_{1\,2} + \beta \, (P_3 \cos\phi + P_t \sin\phi), P_3 \sin\phi - P_t \cos\phi \rangle;
\]

subalgebras of dimension 3:
\[
    \langle P_1, P_3, P_t \rangle, \ \ \langle J_{1\,2}, P_3, P_t \rangle,
\]
\[
    \langle P_1 + \alpha \, (P_3 \cos\phi + P_t \sin\phi), P_2, P_3 \sin\phi - P_t \cos\phi \rangle,
\]
\[
    \langle J_{1\,2} + \beta \, (P_3 \cos\phi + P_t \sin\phi), P_1, P_2 \rangle;
\]

subalgebras of dimension 4:
\[
    \langle P_1, P_2, P_3, P_t \rangle, \ \ \langle J_{1\,2} + \beta \, (P_3 \cos\phi + P_t \sin\phi), P_1, P_2, P_3 \sin\phi - P_t \cos\phi \rangle;
\]

subalgebras of dimension 5:
\[
   \langle J_{1\,2}, P_1, P_2, P_3, P_t \rangle.
\]

Here, $\alpha \geq 0$ and $\beta$ are arbitrary real constants and $0 \leq \phi < \pi$.

As an example, consider the algebra $\langle J_{1\,2}, P_3 \sin\phi - P_t \cos\phi \rangle$. solving the corresponding system of Lagrange equations, we easily obtain the following ansatz for this algebra:
\begin{equation}\label{6}
    u = u(r, z), \ \ v = v(r, z), \ \ S_k = S_k(r, z), \ \ \ k = 1, 2,
\end{equation}
where $z = x_3 - \mu t$ and $r = \sqrt{\vphantom{\int} {x_1^2 + x_2^2}}$ are new independent variables admit an obvious physical interpretation: the first variable performs a transition to a moving coordinate system (along the variable $x_3$) with origin on the surface $S_1 = 0$, and the second variable indicates the radial symmetry of the process with respect to the variables $x_1$ and $x_2$.

We substitute ansatz \eqref{6} into the boundary-value problem \eqref{1}--\eqref{5} with $\boldsymbol{Q}(t) = \boldsymbol{q}$. After the corresponding calculations, we obtain the following boundary-value problem for a two-dimensional system of equations of the elliptic type:
\begin{equation}\label{7}
    \frac{1}{r}\left( r d_1(u) u_r \right)_r + \left( d_1(u) u_z \right)_z + \mu u_z = 0,
\end{equation}
\begin{equation}\label{8}
    \frac{1}{r}\left( r d_2(v) v_r \right)_r + \left( d_2(v) v_z \right)_z + \mu v_z = 0,
\end{equation}
\begin{equation}\label{9}
    S_1(r, z) = 0: \, d_{1\,v}{\nabla}'u \cdot {\nabla}' S_1 = \left( \mu H_v - q \right) \frac{\partial S_1}{\partial z}, \ \ u = u_v,
\end{equation}
\begin{equation}\label{10}
    S_2(r, z) = 0: \, d_{2\,m}{\nabla}'v \cdot {\nabla}' S_2 = d_{1\,m}{\nabla}'u \cdot {\nabla}' S_2 + \mu H_m \frac{\partial S_2}{\partial z}, \ \ u = u_m,  \ \ v = v_m,
\end{equation}
\begin{equation}\label{11}
    r^2 + z^2 = + \infty: \, v = v_{\infty},
\end{equation}
where
\[
    {\nabla}' = \left( \frac{\partial}{\partial r}, \frac{\partial}{\partial z} \right),
\]
$\mu$ is an unknown parameter, and $r$ and $z$ denote differentiation with respect to these variables.

Although the obtained boundary-value problem is much simpler than the original problem \eqref{1}--\eqref{5}, it remains a nonlinear problem with basic two-dimensional partial differential equations. Our main aim is to reduce problem \eqref{7}--\eqref{11} to a boundary-value problem with basic ordinary differential equations. For this purpose, one can use different approaches, but we consider here only one interesting example. Consider the ansatz \cite{iva47}
\begin{equation}\label{12}
    u = u(\omega), \ \ v = v(\omega), \ \ S_k = S_k(\omega), \ \ \omega = z + \sqrt{\vphantom{\sum} {z^2 + r^2}}, \ \ \ k = 1, 2.
\end{equation}
Note that this is a non-Lie ansatz because the maximal invariance algebra of system \eqref{7} and \eqref{8} with arbitrary functions $d_1(u)$ and $d_2(v)$ is trivial and is generated by the shift operator~$\partial / \partial z$.

Substituting ansatz \eqref{12} into the boundary-value problem \eqref{7}--\eqref{11} and performing the corresponding calculations, we obtain the following boundary-value problem for a system of ordinary differential equations:
\begin{equation}\label{13}
    \frac{d}{d \omega} \left( \omega \, d_1(u) \frac{du}{d\omega} \right) + \mu \frac{\omega}{2} \frac{du}{d\omega} = 0,  \ \ 0 < \omega_1 < \omega < \omega_2,
\end{equation}
\begin{equation}\label{14}
    \frac{d}{d \omega} \left( \omega \, d_2(v) \frac{dv}{d\omega} \right) + \mu \frac{\omega}{2} \frac{dv}{d\omega} = 0,  \ \ \omega > \omega_2,
\end{equation}
\begin{equation}\label{15}
    \omega = \omega_1: \, 2 d_{1\,v} \frac{du}{d\omega} = \mu H_v - q, \ \ u = u_v,
\end{equation}
\begin{equation}\label{16}
    \omega = \omega_2: \, 2 d_{2\,m} \frac{dv}{d\omega} = 2 d_{1\,m} \frac{du}{d\omega} + \mu H_m, \ \ u = u_m, \ \ v = v_m,
\end{equation}
\begin{equation}\label{17}
    \omega = + \infty: \, v = v_{\infty},
\end{equation}
where $\omega_k$, $k = 1, 2$, and $\mu$ are unknown parameters that must be determined in the solution of the problem.

We can now determine the form of the free surfaces $S_k(t, \boldsymbol{x}) = 0$, $k = 1, 2$, because, according to ansatz \eqref{12},
\[
    S_k(\omega) \equiv z + \sqrt{\vphantom{\sum} {z^2 + r^2}} = \omega_k, \ \ \ k = 1, 2.
\]
The last equation can be rewritten in the form
\begin{equation}\label{18}
    \frac{\vphantom{\int}{x_1^2 + x_2^2}}{\omega_k^2} = 1 - \frac{2z}{\omega_k}, \ \ \ k = 1, 2.
\end{equation}
Thus, we have obtained the equation for paraboloids of revolution in the space of the variables $x_1$, $x_2$, and $z$. From the physical point of view, the unknown parameters $\omega_1$ and $\omega_2$ must satisfy the inequalities $\omega_2 > \omega_1 > 0$. Moreover, the parameter $\omega_1$ can be determined using the following reasoning: Setting $z = 0$ in Eq. \eqref{18}, we get
\[
    \omega_1 = \sqrt{\vphantom{\sum} {x_1^2 + x_2^2}}.
\]
On the other hand, only the part of the surface $S_1 = 0$ bounded by a circle of radius $R$ receives the flow $\boldsymbol{Q}(t) = \boldsymbol{q}$. Thus, without loss of generality, we can set $\omega_1 = R$.

We now pass to the construction of an exact solution of problem \eqref{13}--\eqref{17}. The main problem here is the solution of the system of nonlinear ordinary differential equations \eqref{13} and \eqref{14} because, in the general case, its general solution is unknown. However, this can be done in some special cases. Note that a solution of problem \eqref{13}--\eqref{17} in the case of linear basic equations (i.e., in the case where $d_1(u) = a_1$ and $d_2(v) = a_2$, $a_1, a_2 \in \mathbb{R}^{+}$) was obtained in \cite{lyu82}. We consider one special case of the nonlinear system of equation \eqref{13} and \eqref{14}, namely the case where $d_1(u) = u^{-1}$ and $d_2(v) = 1$. In this case, its exact solution can be represented in the following implicit form (see, e.g., \cite{zai01}):
\begin{equation}\label{19}
    \int\limits_{a}^{\omega u} \frac{d\nu}{\nu \left( {1 + e^{-W(e^A)+A}} \right)} = \ln\omega + C_2, \ \ v = C_3 \Phi(\omega) + C_4,
\end{equation}
where
\[
    \Phi(\omega) = \int\limits_{\omega}^{+\infty} \omega^{-1} e^{-\mu \omega / 2} d \omega,
\]
$W(x)$ is the Lambert function,
\[
    A = - \frac{\mu}{2} \, \nu + C_1,
\]
$a$ is a certain constant, and $C_1, \ldots, C_4$ are arbitrary constants of integration. Substituting solution \eqref{19} into the boundary conditions \eqref{15}--\eqref{17} and taking into account that
\[
    \frac{d\Phi}{d\omega} = \omega^{-1} e^{-\mu \omega / 2}, \ \ \ln\left( \frac{\omega}{u} \frac{du}{d\omega} \right) + \frac{\omega}{u} \frac{du}{d\omega} = A,
\]
we obtain the required exact solution
\begin{equation}\label{20}
    \int\limits_{R u_v}^{\omega u} \frac{d\nu}{\nu \left( {1 + e^{-W(e^{\mathcal{A}})+\mathcal{A}}} \right)} = \ln\frac{\omega}{R}, \ \ v = \frac{v_m - v_{\infty}}{\Phi(\omega_2)} \, \Phi(\omega) + v_{\infty},
\end{equation}
where the parameters $\omega_2$ and $\mu$ must be determined from the system of transcendental equations
\[
    \int\limits_{R u_v}^{\omega_2 u_m} \frac{d\nu}{\nu \left( {1 + e^{-W(e^{\mathcal{A}})+\mathcal{A}}} \right)} = \ln\frac{\omega_2}{{R}},
\]
\[
    2  \frac{v_m - v_{\infty}}{\Phi(\omega_2)} \, e^{-\mu \omega_2 / 2} = 2 e^{-W\left(e^{\mathcal{A}(\omega_2)}\right)+\mathcal{A}(\omega_2)} + \mu \omega_2 H_m.
\]
Here, we have used the following notation:
\[
    \mathcal{A} = -\frac{\mu}{2} \, \nu + \ln\left( (\mu H_v - q) \frac{R}{2} \right) + (\mu H_v - q) \frac{R}{2} + \frac{\mu}{2} \, R u_v
\]
and
\[
    \mathcal{A}(\omega_2) = \frac{\mu}{2}\,(R u_v - \omega_2 u_m) + \ln\left( (\mu H_v - q) \frac{R}{2} \right) + (\mu H_v - q) \frac{R}{2}.
\]

Using relations \eqref{20}, \eqref{6}, and \eqref{12}, we can now obtain the exact solution of problem \eqref{1}--\eqref{5} for $d_1(u) = u^{-1}$, $d_2(v) = 1$, and $\boldsymbol{Q}(t) = \boldsymbol{q}$ in the implicit form:
\[
    \int\limits_{R u_v}^{\left( \sqrt{\vphantom{\int_a^b}{x_1^2 + x_2^2 + (x_3 - \mu t)^2}} + x_3 - \mu t \right) u} \frac{d\nu}{\nu \left( {1 + e^{-W(e^{\mathcal{A}})+\mathcal{A}}} \right)} = \ln\frac{\sqrt{\vphantom{\int}{x_1^2 + x_2^2 + (x_3 - \mu t)^2}} + x_3 - \mu t}{R},
\]
\[
    v = \frac{v_m - v_{\infty}}{\Phi(\omega_2)} \, \Phi\left( \sqrt{\vphantom{\sum}{x_1^2 + x_2^2 + (x_3 - \mu t)^2}} + x_3 - \mu t \right) + v_{\infty},
\]
\[
    S_k \equiv \frac{\vphantom{\int}{x_1^2 + x_2^2}}{\omega_k^2} + \frac{2(x_3 - \mu t)}{\omega_k} - 1 = 0, \ \ \ k = 1, 2.
\]

\end{document}